\documentclass{PoS}

\usepackage{amsmath}
\usepackage{amsfonts}
\usepackage{amssymb}
\usepackage{amsthm}
\usepackage{graphicx}
\usepackage{subcaption}
\usepackage{graphicx}
\usepackage{mathtools}
\usepackage{float}
\usepackage{bm}  
\usepackage{float}
\usepackage{relsize}
\usepackage{ulem}
\usepackage[shortcuts]{extdash}

\title{Wino dark matter annihilation through the radiative formation of bound states}

\ShortTitle{Wino dark matter annihilation through the radiative formation of bound states}

\author{\speaker{Evan Johnson}
\\
        Department of Physics,
         The Ohio State University, Columbus, OH\ 43210, USA\\
        E-mail: \email{johnson.6036@osu.edu}}

\author{Eric Braaten
\thanks{This work was supported in part by the Department of Energy under grant DE-FG02-05ER15715.}
\\
        Department of Physics,
         The Ohio State University, Columbus, OH\ 43210, USA\\
        E-mail: \email{braaten@mps.ohio-state.edu}}
        
\author{Hong Zhang\\
        Department of Physics,
         The Ohio State University, Columbus, OH\ 43210, USA\\
        E-mail: \email{zhang.5676@osu.edu}}

\abstract{
The most dramatic ``Sommerfeld enhancements'' of neutral-wino-pair annihilation occur when the wino mass is tuned to near critical values where there is a zero-energy S-wave resonance at the neutral-wino-pair threshold.  If the wino mass is larger than the critical value, the resonance is a wino-pair bound state.  If the wino mass is near  a critical value, low-energy winos can be described by a zero-range effective field theory in which the winos interact nonperturbatively through a contact interaction.  The parameters of the zero-range effective field theory can be determined by matching wino scattering amplitudes calculated  by solving the Schr\"odinger equation for a  nonrelativistic effective field theory in which the winos interact through a potential due to the exchange of electroweak gauge bosons.  The utility of the zero-range effective field theory is illustrated by calculating the rate for formation of an S-wave bound state in the collision of two neutral winos through the emission of two soft photons.
}

\FullConference{38th International Conference on High Energy Physics\\
		3-10 August 2016\\
		Chicago, USA}

\emergencystretch=\maxdimen
\begin{document}

\section{Introduction}
A promising way to determine the particle nature of dark matter is by observing its annihilation products through indirect detection experiments. If the dark matter consists of weakly interacting massive particles (wimps),  it is well\=/known that near critical values of the wimp mass, the annihilation rate is greatly enhanced (the so\=/called Sommerfeld enhancements).  For wimp mass near these critical values, there is an S-wave resonance near the wimp\=/pair threshold which provides resonant enhancement of the annihilation rate.  However, such a resonance has other effects.  If the resonance is a bound state, there is a new dark matter annihilation mechanism: formation of the bound state followed by the annihilation of its constituents.  Previous work on the formation of bound states of dark matter particles is summarized in Ref.~\cite{Asadi:2016ybp}.  Much of the previous work focused on wimps coupling to dark sector mediators, such as a light scalar or dark photon.  In this work, as in Ref.~\cite{Asadi:2016ybp}, we focus on wimps with Standard Model electroweak interactions.  Asadi et al.\ calculated the formation rate of P-wave bound states in the collision of a pair of wimps through a radiative transition \cite{Asadi:2016ybp}.  In this work, we describe a zero-range effective field theory for wimps near a critical mass.  It greatly simplifies the calculation of reaction rates for the S-wave bound state near the threshold.  As an illustration, we compute the formation rate of the S-wave bound state in the collision of a pair of wimps through a double radiative transition.

\section{Fundamental Theory}

We take the dark matter particle to be the neutral member of an $SU(2)$ triplet of Majorana fermions with zero hypercharge.  The fundamental theory could simply be an extension of the Standard Model with this one additional multiplet.  It could also be the Minimal Supersymmetric Standard Model (MSSM) in a region of parameter space where the lightest supersymmetric particle is a wino-like neutralino.  In either case, we refer to  the particles in the $SU(2)$ multiplet as {\it winos}.  We denote the neutral wino by $w^0$ and the charged winos by $w^+$ and $w^-$.  To account for the observed relic density of dark matter, its mass $M$ should be in the TeV range.  If the mass splitting $\delta$ for the charged winos arises from one-loop radiative corrections in the MSSM, $\delta$ is approximately 174~MeV for $M$ in the TeV range \cite{Pierce:1996zz}.  We take $\delta=170$~MeV in this work.

Pairs of neutral winos can scatter or annihilate through their electroweak interactions.  The reactions $w^0 w^0 \to \gamma\gamma \;,\; \gamma Z^0$  produce monochromatic gamma ray signals.    The leading order contribution to these cross sections comes from a one\=/loop diagram with a single $W$\=/boson exchange and goes as $v\sigma \sim \alpha^2 \alpha_2^2/m_W^2$ where $\alpha_2$ is the $SU(2)$ coupling.  This exceeds the unitarity bound $v\sigma < 4\pi/v M^2$ for sufficiently large wino mass $M$ \cite{Hisano:2003ec}.  Higher order diagrams must therefore be included.  These diagrams include the ladder diagrams from exchanges of electroweak gauge bosons, with each `rung' of the ladder introducing a factor of $\alpha_2 M/m_W$ \cite{Hisano:2003ec}.  For large wino mass $M$, these ladder diagrams must be summed to all orders.  

A particularly dramatic consequence of the summation of the ladder diagrams is the existence of a zero\=/energy S-wave resonance at the neutral\=/wino\=/pair threshold $2M$ at a sequence of critical values of $M$ \cite{Hisano:2003ec}.  Near these resonances, the annihilation rate of a wino pair into a pair of electroweak gauge bosons is increased by orders of magnitude \cite{Hisano:2003ec}.  If $\delta= 170$~MeV, the first resonance occurs at the critical mass 2.39~TeV and the second one occurs at 9.30~TeV.  Elastic cross sections and annihilation cross sections for wino pairs can be calculated nonperturbatively in the fundamental theory by summing ladder diagrams to all orders.  However the calculations can be greatly facilitated by using a non\=/relativistic effective field theory for the winos.

\section{Nonrelativistic Effective Field Theory}

\begin{figure}
\centering
\begin{minipage}[t]{.48\textwidth}
  \centering
  \includegraphics[width=\linewidth]{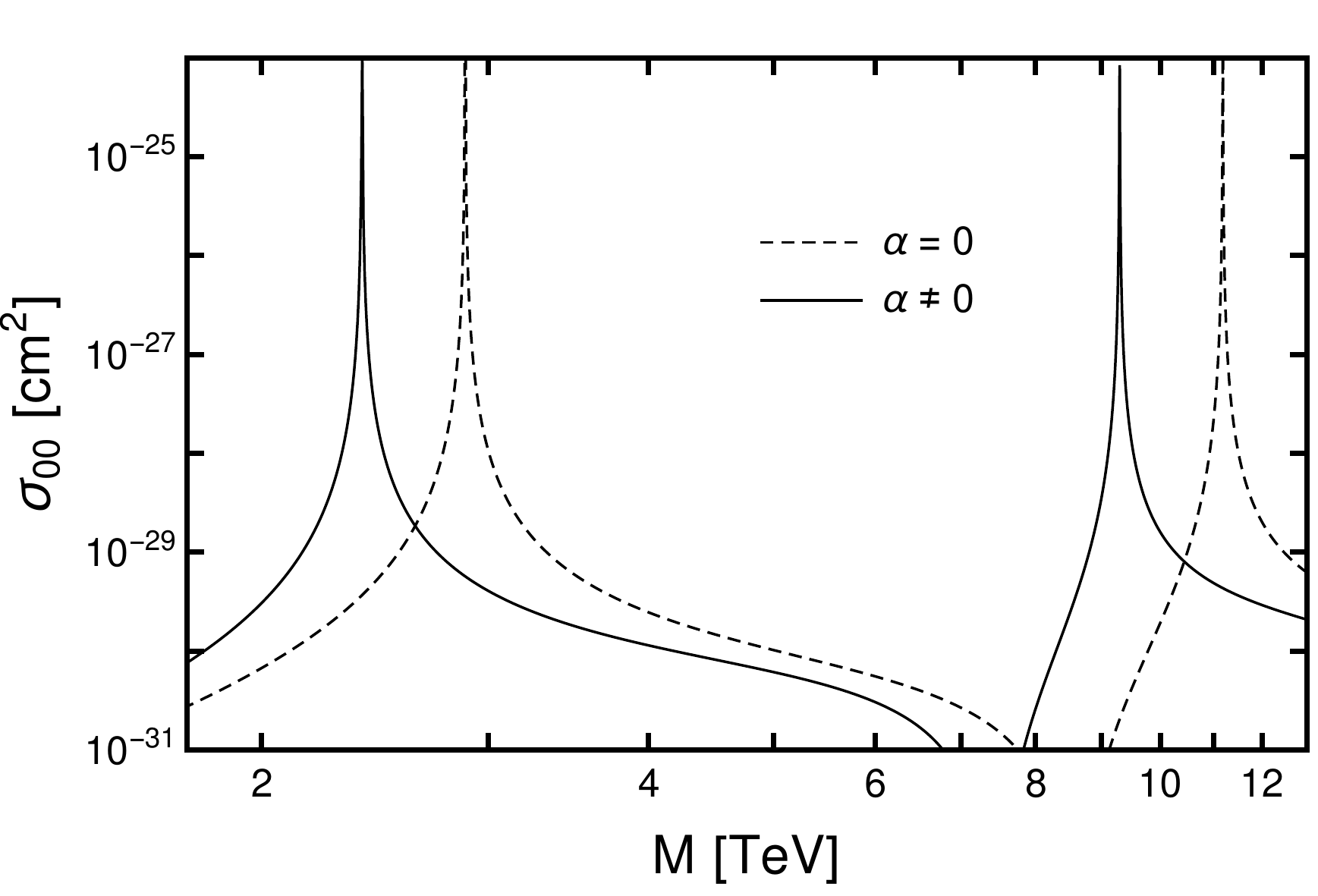}
  \captionof{figure}{The neutral\=/wino elastic cross section $\sigma_{00}$ at zero energy 
  as a function of the wino mass $M$.  Wino pair annihilation reactions are neglected.  The cross section is shown with the Coulomb potential included in $V(r)$ (solid line) and with the Coulomb potential omitted (dashed line).}
  \label{fig:sigma00vsMcode}
\end{minipage}%
\quad
\begin{minipage}[c]{.48\textwidth}
  \centering
\includegraphics[width=0.46\linewidth]{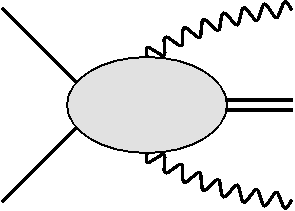}
\quad
\includegraphics[width=0.46\linewidth]{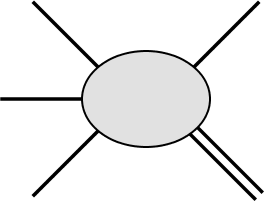}
\vspace*{11mm}
  \captionof{figure}{Wino bound state formation processes: double radiative transition (left panel) and three\=/body recombination (right panel).  Single solid lines represent winos, double solid lines represent a wino\=/pair bound state, and wiggly lines represent photons.}
  \label{fig:BSproduction}
\end{minipage}%
\end{figure}

Nonrelativistic winos can be described by a nonrelativistic effective field theory that we call NREFT in which the winos interact through potentials that arise from the exchange of electroweak gauge bosons \cite{Hisano:2003ec}.  The Schr\"odinger equation with the appropriate potential describing the gauge boson exchanges can be used to sum ladder diagrams to all orders.  The potential matrix for the coupled channels consisting of a pair of neutral winos $w^0 w^0$ and a pair of charged winos $w^+ w^-$ is 
\begin{equation}
V(r) \;=\; \begin{pmatrix} 0 &\; & -\sqrt{2}\alpha_2 \,\frac{e^{-m_W r}}{r} \\ 
-\sqrt{2}\alpha_2 \,\frac{e^{-m_W r}}{r} &\;& 2\delta- 
\frac{\alpha}{r} - \alpha_2 c^2_W\, \frac{e^{-m_Z r}}{r} \end{pmatrix} \;\;,
\label{eq:V-matrix}
\end{equation}
where $c_W=\cos\theta_W$.  The Schr\"odinger equation for the coupled channels can be solved numerically.  It is convenient to choose the zero of energy to be the neutral-wino-pair threshold.  The neutral\=/wino elastic cross section at zero energy is shown as a function of the wino mass in Fig.~\ref{fig:sigma00vsMcode}.

The first two in the sequence of critical masses can be seen in Fig.~\ref{fig:sigma00vsMcode}.  Each peak in the cross-section in Fig.\ref{fig:sigma00vsMcode} corresponds to a critical mass  where there is a zero\=/energy resonance near the $w^0 w^0$ threshold.  With the Coulomb potential included, the first critical mass occurs at 2.39~TeV. When the Coulomb potential is omitted, the value shifts upward to $M_*=2.88$~TeV.  Without the extra attraction provided by the Coulomb potential, a larger mass is needed to form the resonance.  For wino mass $M$ above the critical mass, the resonance is a bound state. Since the bound state energy is negative, its formation in the collision of a pair of low-energy winos requires the emission of photons.
By parity conservation, the emission of a single photon cannot produce an S-wave bound state, but it can produce P-wave bound states.  This radiative transition mechanism for formation of the bound state has been studied in Ref.~\cite{Asadi:2016ybp}.  An S-wave bound state can be formed by the emission of two photons.  This double-radiative transition mechanism is shown schematically in the left panel of Fig.~\ref{fig:BSproduction}.   

For many purposes, the electromagnetic interactions can be treated as a perturbation.  If the Coulomb potential is omitted by setting $\alpha = 0$ in the potential matrix in Eq.~\eqref{eq:V-matrix}, the remaining potentials have a short range of order $1/m_W$.  As a consequence, the inverse of the transition amplitude $\mathcal{T}_{00}$ for elastic neutral\=/wino scattering has an expansion in integer powers of the energy $E$:
\begin{equation}
\frac{8\pi/M}{\mathcal{T}_{00}(E)} =
-\frac{1}{a_0(M)}  - i\sqrt{ME} + \frac{1}{2}r_s(M) ME   + {\cal O}(E^2) \;\;,
\label{eq:T00NRinv}
\end{equation}
where $a_0(M)$ is the {\it S\=/wave scattering length} for neutral winos and $r_s(M)$ is their {\it effective range}.  The inverse scattering length, $1/a_0(M)$, vanishes at the critical mass, $M_*=2.88$~TeV.  The effective range at the critical mass is $r_s(M_*) = -0.692/\sqrt{2M_*\delta}$.  

In addition to using the Schr\"odinger equation for NREFT to calculate the cross sections for scattering of pairs of winos and the rates for their annihilation into electroweak gauge bosons, it can also be used to calculate radiative transition rates of a pair of winos to a bound state by the emission of a soft photon \cite{Asadi:2016ybp}.  If the wino mass is above a critical value but close enough that there is an S\=/wave bound state near the threshold, the calculation of reaction rates involving that S\=/wave bound state can be greatly facilitated by using a zero\=/range effective field theory for the winos.

\section{Zero-Range Effective Field Theory}

Winos with momenta small compared to $m_W$ can be described by a nonrelativistic effective field theory in which winos interact through contact interactions.  We refer to the resulting theory as Zero-Range Effective Field Theory (ZREFT).  If we neglect wino\=/pair annihilation processes, the Lagrangian is
\begin{equation}
\mathcal{L} = w_0^\dagger\left(i\partial_0+\frac{\bm{\nabla}^2}{2M}\right) w_0  
+ \sum_\pm w_\pm^\dagger\left(iD_0+\frac{\bm{D}^2}{2M}-\delta\right) w_\pm +\frac{1}{2}\big(\bm{E}^2-\bm{B}^2\big) +\mathcal{L}_{\rm zr} \;\;,
\label{eq:kineticL}
\end{equation}
where $D_0$ and $\bm{D}$ are the electromagnetic covariant derivatives and $\bm{E}$ and $\bm{B}$ are the electric and magnetic fields. The zero\=/range interaction terms in ${\cal L}_{\rm zr}$ are products of $w_0^a w_0^b$ or $w_+^a w_-^b$ and their hermitian conjugates contracted with the spin\=/singlet projector $\tfrac{1}{2}(\delta^{ac}\delta^{bd}-\delta^{ad}\delta^{bc})$.

\begin{figure}
\centering
\begin{minipage}[t]{.48\textwidth}
  \centering
  \includegraphics[width=\linewidth]{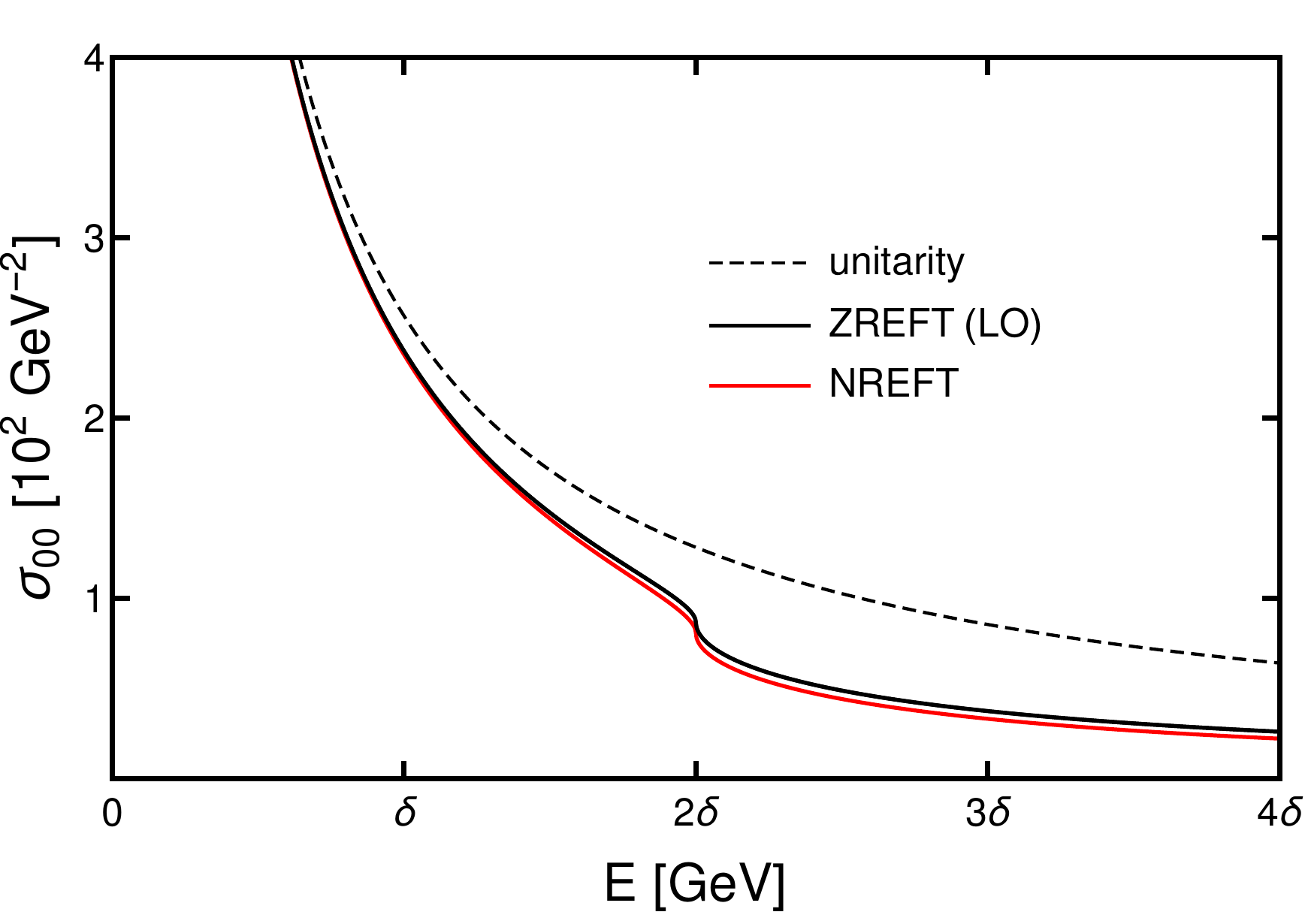}
  \captionof{figure}{The neutral\=/wino elastic cross section $\sigma_{00}(E)$ at the critical wino mass as a function of the energy $E$.  The cross section for $M_*=2.88$~TeV and $\alpha=0$ is shown for NREFT (lower solid red curve)  and for ZREFT at LO (higher solid black curve).  The S\=/wave unitarity bound (dashed line) is saturated at low energy.}
  \label{fig:sigma00vsE-LO}
\end{minipage}%
\quad
\begin{minipage}[t]{.48\textwidth}
  \centering
  \includegraphics[width=\linewidth]{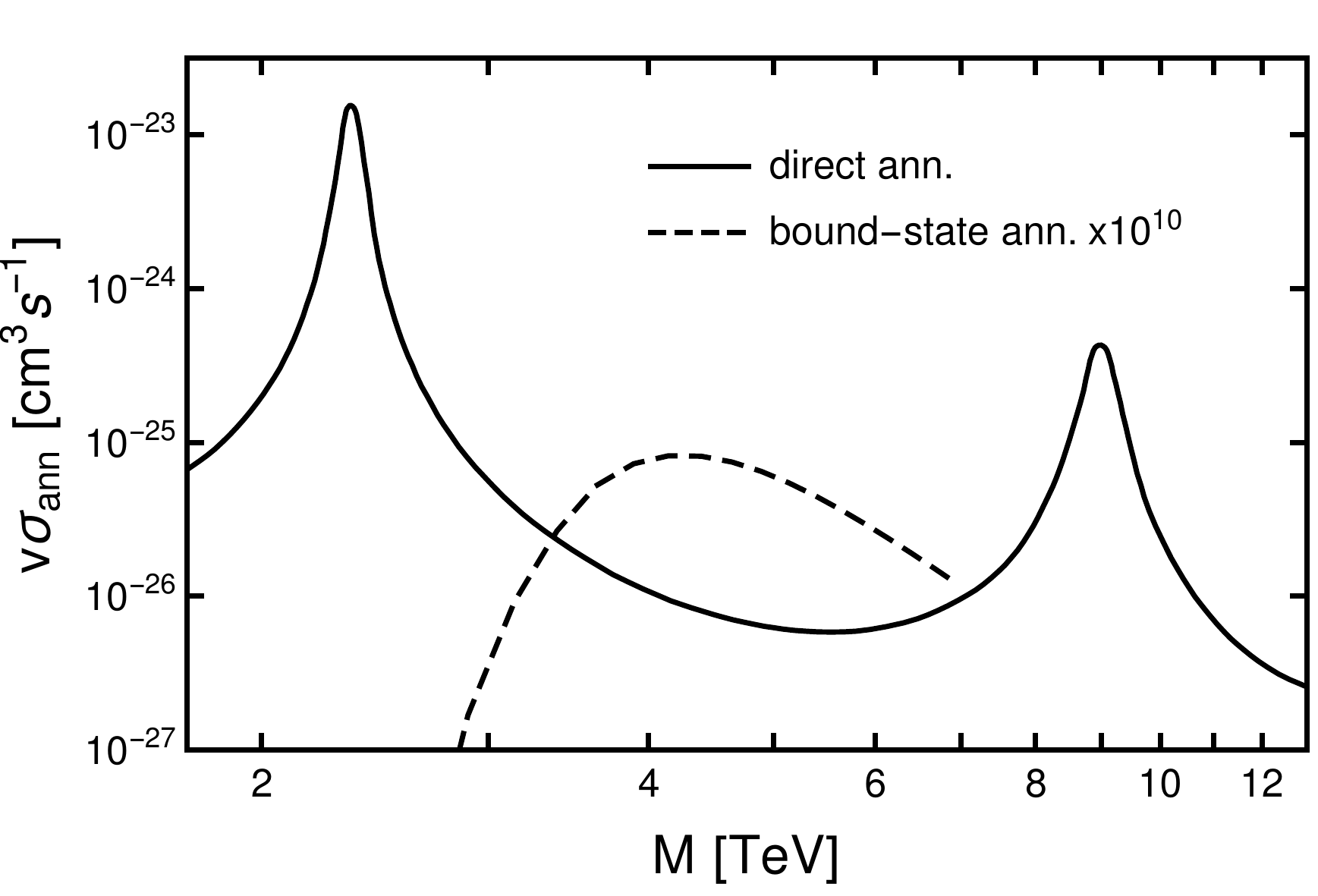}
  \captionof{figure}{Annihilation rates $v\sigma_{\mathrm{ann}}$ for neutral winos with velocity $v = 10^{-3}$ in the center-of-mass frame into monochromatic gamma rays as a function of the wino mass $M$: direct annihilation rate (solid curve) and annihilation rate of bound states formed by a double radiative transition (dashed curve).}
  \label{fig:BS-rate}
\end{minipage}
\end{figure}

Lensky and Birse have carried out a careful renormalization group (RG) analysis of the two\=/particle sector for the field theory with two coupled scattering channels with zero\=/range interactions \cite{Lensky:2011he}. They identified three distinct RG fixed points, corresponding to the number of fine\=/tuned parameters.  In the present context, only the wino mass $M$ is tuned to be near a critical value.  The appropriate fixed point corresponds to scattering that saturates the S\=/wave unitarity bound in a channel that is a linear combination of $w^0 w^0$ and $w^+ w^-$ with a mixing angle $\phi$ and no scattering in the orthogonal channel.  At leading order in the power counting, the transition amplitude for neutral\=/wino scattering below the $w^+ w^-$ threshold is
\begin{equation}
{\cal T}_{00}(E) = \frac{8\pi /M}{-1/a_0(M) + \tan^2\phi\, (\sqrt{\Delta^2 - ME}-\Delta)\,  - i \, \sqrt{ME}} \;\;,
\label{eq:T00LOloEu}
\end{equation}
where $\Delta=\sqrt{2M\delta}$ and $a_0(M)$ is the S\=/wave scattering length.  The scattering length can be calculated as a function of $M$ by solving the Schr\"odinger equation numerically in NREFT. Its inverse $1/a_0(M)$ vanishes at the critical mass $M_*$.  The effective range in ZREFT at LO is predicted by Eq.~\eqref{eq:T00LOloEu} to be $r_s(M) = -\tan^2\phi/\Delta$.  To fix the angle $\phi$, we choose to match the effective range at the critical mass, which gives $\tan\phi=0.832$.  We illustrate the effectiveness of ZREFT in Fig.~\ref{fig:sigma00vsE-LO} by comparing the neutral\=/wino elastic cross section calculated numerically by solving the Schr\"odinger equation with the potential in Eq.~\eqref{eq:V-matrix} with $\alpha=0$ with that calculated in the ZREFT at LO with $\tan\phi=0.832$.  The analytic result from ZREFT at LO reproduces the cross section calculated numerically in NREFT with surprising accuracy, including the non\=/trivial behavior at the $w^+ w^-$ threshold at $2\delta=0.34$~GeV.

\begin{figure}
\centering
\includegraphics[width=\linewidth]{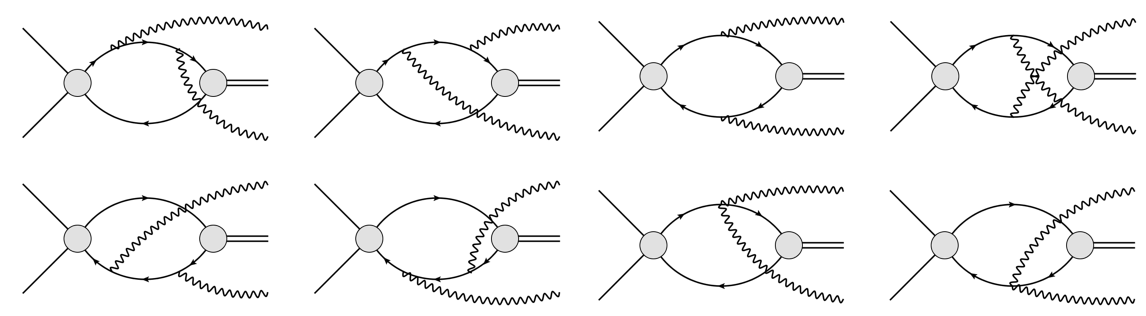}
\caption{Diagrams that contribute to the bound\=/state formation process $w^0 w^0 \to (ww) + \gamma\gamma$.  In each diagram, the left grey blob represents the $w^0 w^0$ to $w^+ w^-$ transition amplitude and the right blob represents the $w^+ w^-$ to $(ww)$ transition amplitude.}
\label{fig:BS_diagrams}
\end{figure}

Once the parameters of ZREFT are fixed, the effective field theory can be used to calculate the rates for the bound\=/state formation processes in Fig.~\ref{fig:BSproduction}.  For the double radiative transition process, the contributing diagrams are shown in Fig.~\ref{fig:BS_diagrams}.  The matrix element can be calculated analytically as a function of the collision energy and the two photon energies.  
The matrix element is particularly simple if $M$ is close enough to $M_*$ that the scattering length satisfies $1/Ma_0^2 \ll \delta$.  In the scaling region of the wino velocity defined by $1/Ma_0 \ll v \ll (\delta/M)^{1/2}$, 
the double\=/radiative transition rate is
\begin{equation}
v\sigma \approx \frac{\tan^2\!\phi\,\alpha^2 \,M^2 \,\hbar^3}{26880\,a_0 \,\delta^5\, c^2}\; (E/Mc^2)^{6}  \;\; ,
\end{equation}
where $E=M v^2/4$ is the total energy of the colliding winos in the center\=/of\=/mass frame.  Once the bound state is produced, it will annihilate into electroweak gauge bosons, with probability 1.  Its annihilation into $\gamma \gamma$ and $\gamma Z^0$ provides an additional mechanism for producing monochromatic gamma rays, complimentary to direct wino annihilation.  This contribution to the annihilation rate is equal to the formation rate.  We compare it to the direct wino\=/pair annihilation rate in Fig.~\ref{fig:BS-rate}.  Near the resonance, the bound-state formation rate scales as $v^{12}$, compared to $v^{-2}$ for the direct annihilation rate, leading to a high suppression for small $v$.  As shown in Fig.~\ref{fig:BS-rate}, the ratio of the bound-state formation rate to the direct annihilation rate is always smaller than $10^{-9}$.

ZREFT can also be applied to the formation of the bound state by the three\=/body recombination process (right panel of Fig.~\ref{fig:BSproduction}).  The three\=/body recombination rate can be calculated by solving a single\=/variable integral equation \cite{Braaten:2004rn}.  This calculation is much easier than solving the three\=/body Schr\"odinger equation of NREFT.  The three\=/body recombination rate is proportional to $n^3$, where $n$ is the wino number density, compared to $n^2$ for the direct annihilation rate.  Thus it becomes increasingly important near the center of a dark matter halo.  In the scaling region of the wino velocity, the coefficient of $n^3$ in the three\=/body recombination rate scales as $1/v^4$.  Thus three\=/body recombination may be particularly important in dwarf galaxies where $v$ can be very small.


\begin{thebibliography}{99}

\bibitem{Asadi:2016ybp} 
  P.~Asadi, M.~Baumgart, P.~J.~Fitzpatrick, E.~Krupczak and T.~R.~Slatyer,
  ``Capture and decay of electroweak WIMPonium,''
  [arXiv:1610.07617 [hep-ph]].

\bibitem{Pierce:1996zz} 
  D.~M.~Pierce, J.~A.~Bagger, K.~T.~Matchev and R.~J.~Zhang,
  ``Precision corrections in the minimal supersymmetric standard model,''
  Nucl.\ Phys.\ B {\bf 491}, 3 (1997)
  [hep-ph/9606211].

\bibitem{Hisano:2003ec} 
  J.~Hisano, S.~Matsumoto and M.~M.~Nojiri,
  ``Explosive dark matter annihilation,''
  Phys.\ Rev.\ Lett.\  {\bf 92}, 031303 (2004)
  [hep-ph/0307216].


\bibitem{Lensky:2011he} 
  V.~Lensky and M.~C.~Birse,
  ``Coupled-channel effective field theory and proton-$^7$Li scattering,''
  Eur.\ Phys.\ J.\ A {\bf 47}, 142 (2011)
  [arXiv:1109.2797 [nucl-th]].

\bibitem{Braaten:2004rn} 
  E.~Braaten and H.-W.~Hammer,
``Universality in few-body systems with large scattering length,''
  Phys.\ Rept.\  {\bf 428}, 259 (2006)
  [cond-mat/0410417].


\end{thebibliography}
\end{document}